\documentclass[aps,prb,twocolumn,groupedaddress,showpacs,intlimits,amsmath,amssymb,floatfix,citeautoscript,footinbib]{revtex4-1}
\usepackage{bm}
\usepackage[final]{graphicx}
\usepackage[hang,center]{subfigure}

\emergencystretch=\maxdimen
\begin{document}

\title{Quantifying electronic correlation strength in a  complex oxide: a combined DMFT and ARPES study of LaNiO$_3$}

\author{E. A. Nowadnick$^{1,2}$}
\email{E-mail: nowadnick@cornell.edu}
\author{J. P. Ruf$^{3}$}
\author{H. Park$^{2,4,5,6}$}
\author{P. D. C. King$^{3,7}$}
\altaffiliation{ Current address: SUPA, School of Physics and Astronomy, University of St. Andrews, St. Andrews KY16 9SS, UK}
\author{D. G. Schlom$^{7,8}$}
\author{K. M. Shen$^{3,7}$}
\author{A. J. Millis$^{2}$}
\affiliation{$^1$School of Applied and Engineering Physics, Cornell University, Ithaca, NY 14853, USA}
\affiliation{$^2$Department of Physics, Columbia University, New York, NY 10027, USA}
\affiliation{$^3$Department of Physics, Cornell University, Ithaca, NY 14853, USA}
\affiliation{$^4$Department of Applied Physics and Applied Mathematics, Columbia University, New York, NY 10027, USA}
\affiliation{$^5$ Department of Physics, University of Illinois at Chicago, Chicago, IL 60607, USA}
\affiliation{$^6$ Materials Science Division, Argonne National Laboratory, Argonne, IL 60439, USA}
\affiliation{$^7$Kavli Institute at Cornell for Nanoscale Science, Ithaca, NY 14853, USA}
\affiliation{$^8$ Department of Materials Science and Engineering, Cornell University, Ithaca, NY 14853, USA}

\date{\today}

\begin{abstract}
The electronic correlation strength is a basic quantity that characterizes the physical properties of materials such as transition metal oxides. Determining correlation strengths requires both precise definitions and a careful comparison between experiment and theory. In this paper we define the correlation strength via the magnitude of the electron self-energy near the Fermi level. For the case of LaNiO$_3$, we obtain both the experimental and theoretical mass enhancements $m^\star/m$ by considering high resolution angle-resolved photoemission spectroscopy (ARPES) measurements and density functional + dynamical mean field theory (DFT + DMFT) calculations. We use valence-band photoemission data to constrain the free parameters in the theory, and demonstrate a quantitative agreement between the experiment and theory when both the realistic crystal structure and strong electronic correlations are taken into account. In addition, by considering DFT + DMFT calculations on epitaxially strained LaNiO$_3$, we find a strain-induced evolution of $m^\star/m$ in qualitative agreement with trends derived from optics experiments. These results provide a benchmark for the accuracy of the DFT+DMFT theoretical approach, and can serve as a test case when considering other complex materials. By establishing the level of accuracy of the theory, this work also will enable better quantitative predictions when engineering new emergent properties in nickelate heterostructures. 
\end{abstract}


\maketitle

\section{Introduction}
Strongly correlated electron materials such as transition metal oxides display a rich variety of phenomena, including superconductivity, magnetism, and charge and orbital orders.~\cite{imada} Electron-electron interactions (among the electrons in partially filled transition metal $d$-orbitals in the transition metal oxide case) are a key source of this richness, but other factors  including the crystal structure~\cite{Pavarini2004} and the relative energies of other orbitals~\cite{Zaanen85} also are important. 
A defining property of a  strongly correlated electron material is the ``correlation strength''  which characterizes  the degree to which a measured property of the material differs from that predicted by a reference calculation in which electron-electron interactions are neglected or treated in a mean-field manner.  
While the qualitative meaning of the term ``correlation strength'' is intuitively clear, a  precise definition of correlation strength  requires  choices of a reference calculation and a physical property. The most widely used reference calculation is density functional theory (DFT) while the most basic  property is the electron mass enhancement $m^\star/m$, which can be accessed by a variety of experimental probes. An important test of a theoretical method is whether it reproduces the correlation strength of a given material.

In this paper we investigate the mass enhancement of LaNiO$_3$ both experimentally, via angle-resolved photoemission spectroscopy (ARPES)~\cite{Damascelli2003}  and theoretically, via  density functional + dynamical mean field  theory (DFT + DMFT) calculations. By comparing the experiment and theory at precisely the same points in momentum space, and to the same reference DFT bandstructure, we seek to quantify how well DFT + DMFT can reproduce the experimental mass enhancement.  LaNiO$_3$ is an appropriate test case for this study because it has strong electronic correlations, yet remains a Fermi liquid down to low temperatures.   In addition, there has been a great deal of interest in potential exotic physics in LaNiO$_3$, including  engineering orbital polarization with epitaxial strain~\cite{Wu2013,peil,He2015} and a cuprate-like Fermi surface via heterostructuring,~\cite{hansmann} and a precise knowledge of the predictive capabilities of  theory will aid in engineering new emergent properties in the nickelates.

LaNiO$_3$ is also an interesting choice for this study because there is a strong association between correlation effects such as metal-insulator transitions (MIT) and  lattice distortions in the nickelates. Considering the phase diagram of the rare-earth nickelate family $R$NiO$_3$, LaNiO$_3$  is the only material that remains metallic to lowest temperatures; upon cooling other   $R$NiO$_3$ undergo a MIT accompanied by a large-amplitude structural distortion.\cite{torrance} The MIT temperature  is related to the amplitude of the NiO$_6$ octahedral rotations, which in bulk materials is controlled by the size of the $R$ cation.  In addition, studies of ultrathin films of LaNiO$_3$ grown epitaxially on different substrates ~\cite{boris,gray,scherwitzl,son,king,yoo}  show that both the critical film thickness for a MIT and the coefficient of the $T^2$ term in the resistivity depend on the value of the epitaxially-induced strain. We thus expect that in LaNiO$_3$ the correlation strength is determined by a subtle interplay between a basic interaction parameter and the precise crystal structure. A precise determination of the correlation strength must untangle these two effects; this issue is relevant for many transition metal oxides. 

LaNiO$_3$ has already been characterized experimentally via ARPES~\cite{Eguchi,king,yoo}, optical conductivity,~\cite{Ouellette, stewart,stewart2} and thermodynamic measurements,~\cite{xu,sreedhar,rajeev}  as well as theoretically with DFT~\cite{hamada,chakhalian,may}, DFT+DMFT~\cite{Deng,park,park2,han,hansmann,peil} and model system~\cite{lee,johnston,lau} calculations, and a variety of measurements of $m^\star/m$ have previously appeared (to be discussed more later).
ARPES offers a particularly direct measure of a material's mass enhancement via momentum-resolved access to the electronic Green's function, which is the fundamental quantity that characterizes electron propagation in a material: 
\begin{equation}
{G}({\bf k},\omega)=\left(\omega+\mu-{H}_{DFT}({\bf k})-{\Sigma}({\bf k},\omega)\right)^{-1}.
\label{Gdef}
\end{equation}
 Here ${\bf k}$ is a momentum in the first Brillouin zone, $\mu$ is the chemical potential and ${H}_{DFT}$ is the reference calculation Hamiltonian (here labelled by DFT because this is the reference used in this paper).  The electron self-energy ${\Sigma}({\bf k},\omega)$ encodes the electronic correlations by  parametrizing the difference between the observed electron propagation and that predicted  by ${H}_{DFT}({\bf k})$. Below we briefly summarize the relevant theory for extracting a mass enhancement of a Fermi liquid from ARPES.

In a Fermi liquid at the lowest temperatures and frequencies and small (${\bf k}-{\bf k}_F$), the imaginary part  of the self-energy $\Sigma''({\bf k},\omega)$ may be neglected. In this case, the physical Fermi surface is given by the locus of momentum points ${\bf k}={\bf k}_F$  for which $\mathrm{det}\left[{H}_{DFT}({\bf k}_F)+\Sigma'({\bf k}_F, \omega=0)\right]=\mu$, while the DFT Fermi surface is defined as points  ${\bf k}={\bf k}_F$  for which $\mathrm{det}\left[{H}_{DFT}({\bf k}_F)\right]=\mu_{DFT}$, where $\mu$ and  $\mu_{DFT}$ are chemical potentials chosen so that the electron density in each calculation is the same as that in the actual material. Remarkably, in many cases including that studied here,  DFT correctly predicts the {\em shape} of the Fermi surface; correlations are revealed only in the excitation spectrum. Of particular interest is the physical quasiparticle dispersion $\omega_{qp}({\bf k})$ of the correlated material, given by the solution of 
\begin{equation}
\mathrm{det}\left[\omega_{qp}({\bf k})-H_{DFT}({\bf k}) - \Sigma'({\bf k}, \omega_{qp}({\bf k}))+\mu\right]=0.
\label{qpdispersion}
\end{equation}
While Eq. ~\ref{qpdispersion} is defined for all ${\bf k}$ the solution describes a propagating Fermi liquid quasiparticle only for ${\bf k}$ near ${\bf k}_F$.  In this range of momenta, comparison of the  quasiparticle dispersion to the DFT  dispersion $\omega_{DFT}({\bf k})$ defined by
\begin{equation}
\mathrm{det}\left[\omega_{DFT}({\bf k})-H_{DFT}({\bf k}) +\mu_{DFT}\right]=0
\end{equation}
gives a mass enhancement $m^\star/m$ relative to DFT. This mass enhancement is widely regarded as a key measure  of correlation strength. The mass enhancement can in principle be momentum-dependent, either through intrinsic momentum dependence of the electron self-energy $\Sigma({\bf k},\omega)$, which is ignored in the single-site DMFT approximation, or if the orbital character varies substantially around the Fermi surface (for example if the Fermi surface has regions of both predominantly strongly correlated $d$ and weakly correlated $p$ character). 

The plan for the rest of the paper is as follows. We describe our methods in Sec.~\ref{sec:methods}, and then consider the choice of the free parameters in the DFT + DMFT calculation (via comparison to high-energy features in the experimental angle-integrated photoemission spectra) and the choice of the reference DFT bandstructure in Sec.~\ref{sec:ref}. We present the results of our study in Sec.~\ref{sec:results}, and discuss our results in the context of the literature and conclude in Secs.~\ref{sec:lit} and ~\ref{sec:conclusion}.

\section{Methods \label{sec:methods}}
Epitaxial thin films of LaNiO$_3$ are grown on (001) oriented pseudocubic ($pc$) LaAlO$_3$ substrates using reactive-oxide molecular-beam epitaxy and then transferred and measured under ultra-high vacuum with $in$ $situ$ ARPES via the methods described in Ref.~\onlinecite{king}. All ARPES data reported here were obtained with a VG Scienta R4000 electron analyzer using He I$\alpha$ radiation ($h\nu = 21.2$ eV) at a measurement temperature of $T = 20$ K and with 8 meV energy resolution. Because we are interested in the correlated-metal properties here, we use films with a 10 pseudocubic unit cell thickness where a bulk-like Fermi surface has been observed, well away from the previously reported thickness-driven MIT.~\cite{king} 

DFT calculations are performed using the Perdew-Burke-Ernzerhof (PBE) functional and the projector augmented wave method as implemented in the Vienna Ab-initio Simulation Package (VASP).~\cite{kresse1996,kresse1996_2} We use a 600 eV plane wave cutoff, and for structural relaxations a force convergence tolerance of 2 meV/$\AA$.  We consider bulk LaNiO$_3$ in both the rhombohedral (space group $R\bar{3}c$, $a^-a^-a^-$ in Glazer notation) and idealized cubic (space group $Pm\bar{3}m$) structures. For the rhombohedral and cubic structures we use $7\times 7\times 7$ and $8\times 8 \times 8$ $k$-point meshes, respectively. We also consider LaNiO$_3$ under biaxial strain, in which case the symmetry is reduced from rhombohedral to monoclinic (space group $C2/c$,  in Glazer notation $a^-a^-c^-$). For biaxial strain calculations we use a 10 atom unit cell with lattice vectors ($a$, $a$, 0), ($a+\Delta$, $\Delta$, $c$) and ($\Delta$, $a+\Delta$, $c$). This choice of  unit cell imposes an epitaxial constraint to a square substrate with in-plane pseudocubic lattice constant $a$. This cell allows for relaxation of the out-of-plane lattice parameter $c$ as well as  a monoclinic tilt $\beta$ of the unit cell ($\tan \beta = c/\Delta$). We relax $\beta$ by manually setting $\Delta$ to different values, relaxing the resulting structure, and choosing the value of $\Delta$ that yields the minimum total energy.  We make use of VESTA~\cite{momma2008vesta} to visualize crystal structures.   

DFT + DMFT calculations are performed using the methodology described in Ref.~\onlinecite{park2014} using structures relaxed within DFT. DFT calculations are fit using Wannier90~\cite{mostofi2008} over the full $\approx$ 10 eV range spanned by the $p$-$d$ manifold to obtain the correlated Ni-$d$ subspace. Interactions in the correlated subspace are taken to be of the Slater-Kanamori form specified by the Hubbard interaction strength $U$ and the  Hund's couplings $J$. As described below, the values of these parameters are fixed by comparison to wide energy range angle-integrated photoemission measurements. For the double-counting correction required in DFT + DMFT~\cite{wang} we use the parameterization $U^\prime$ = $U-0.2$ eV, which was found to correctly reproduce the pressure phase diagram of the $R$NiO$_3$ family.~\cite{park2} The filled $t_{2g}$ orbitals are treated with the Hartree-Fock approximation while the partially occupied $e_g$ orbitals are treated with single-site DMFT. The DMFT impurity problem is solved using the hybridization expansion version of continuous time quantum Monte Carlo~\cite{werner06,haule} with temperature set  to 0.01 eV $~\approx 120$ K.   For analytic continuation to obtain the real frequency spectral function and density of states (DOS), we employ the Maximum Entropy Method.~\cite{jarrell1996}

\section{Determination of interaction strength and reference bandstructure \label{sec:ref}}
In making a quantitative comparison between experiment and theory, it is important to clarify the relevant uncertainties. For our  comparison of the  experimental and theoretical $m^\star/m$ for LaNiO$_3$, the uncertainties arise from the free parameters  in the DFT + DMFT calculation, and the choice of reference DFT bandstructure. We explore both of these issues in turn in this section.

\subsection{Interaction parameters}

\begin{figure}
\includegraphics[width=0.4\textwidth]{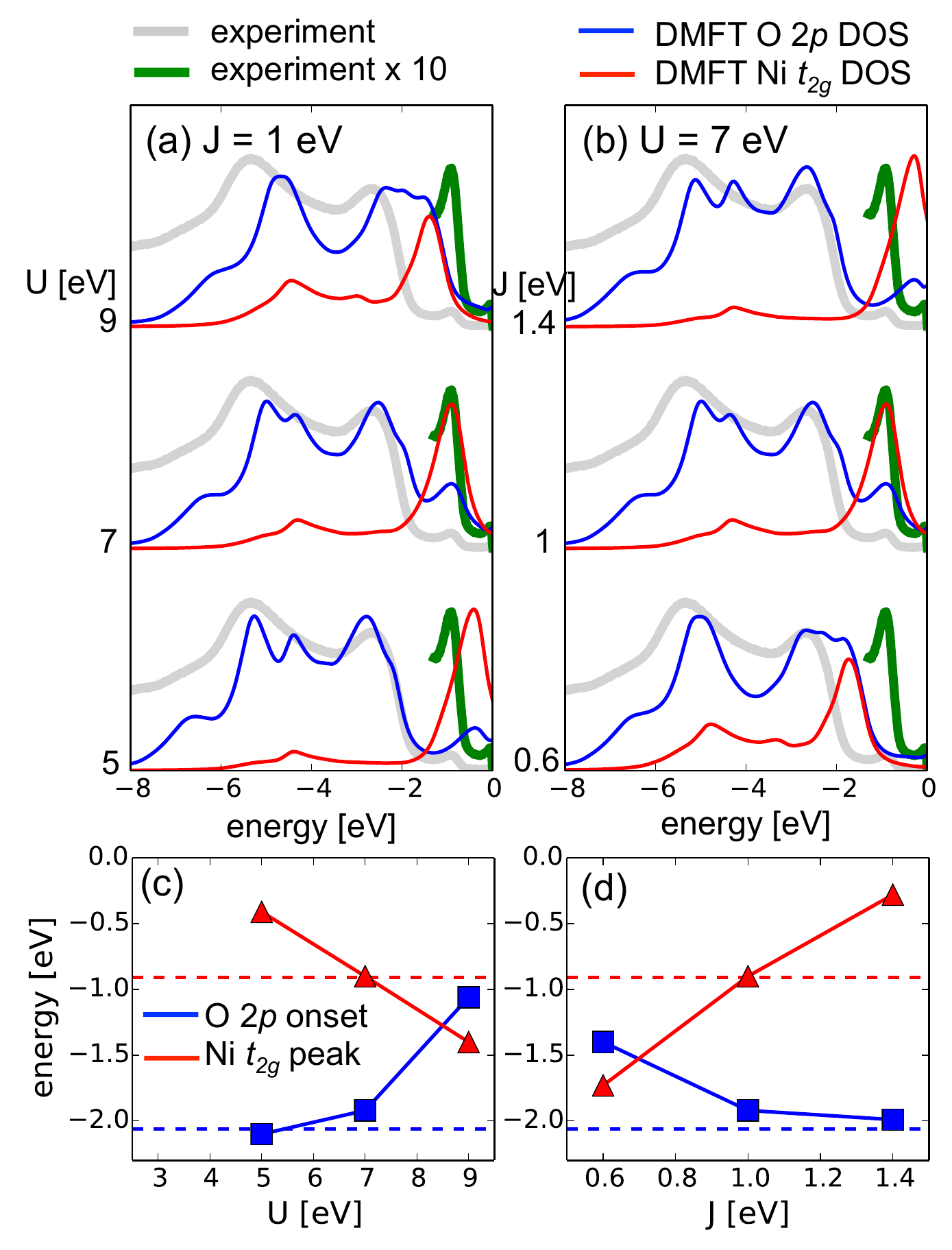}
\caption{\label{fig:1} Comparison of experimental angle-integrated photoemission spectra (thick lines, grey and green online) to calculated DFT + DMFT partial density of states (thin lines, blue and red online) for (a) $J$ = 1 eV and $U$ = 5, 7, and 9 eV (bottom to top), and (b) $U$ = 7 eV and $J$ = 0.6, 1.0, and 1.4 eV (bottom to top). Due to the experimental photon energy, the intensity of the Ni-$t_{2g}$ peak is weak compared to the O-2$p$ features, so the low energy part of the spectrum enlarged by a factor of 10 (green) is shown for comparison to the calculated Ni-$t_{2g}$ peak.  Note that the spectra are artificially offset along the vertical axis for clarity.
(c) and (d) plot the midpoint of the O-$2p$ onset (blue squares) and the Ni-$t_{2g}$ peak position (red triangles) extracted from the DFT + DMFT calculations in (a) and (b) as a function of $U$ and $J$ respectively. The  dashed lines indicate the position of these features in the experimental spectrum. 
}
\end{figure}

We first fix the  Hubbard $U$ and Hund's interaction $J$ interaction parameters by comparing experimental angle-integrated photoemission spectra to the DFT + DMFT DOS  calculated with various choices of ($U$, $J$) in Fig.~\ref{fig:1}. The DFT + DMFT calculations  are performed using a $R\bar{3}c$ crystal structure relaxed within DFT. The experimental spectra  show a  peak at $\sim 1$ eV binding energy, arising from the Ni $t_{2g}$ states,   and a broad higher-energy feature with onset  at $\sim2$ eV binding energy, arising from the O 2$p$ states. We note that the calculation does not include matrix element effects, which  are crucial in determining photoemission intensities.  As a result, in comparing the measurement and the calculation, we focus on peak positions and the onset of spectral weight, rather than the precise shape and size of the peaks.

Fig.~\ref{fig:1}(a) compares our experimental spectrum to DFT + DMFT DOS calculations with different values of $U$ (5, 7, and 9 eV), while keeping $J$ fixed to 1 eV.  From the DOS in Fig.~\ref{fig:1}(a) we extract the onset of the O-2$p$ feature (here we define this as the midpoint of the rising edge) and the location of the Ni-$t_{2g}$ peak and plot them in Fig.~\ref{fig:1}(c) as a function of $U$.  Both the $U$ = 5 eV and $U$ = 7 eV calculations capture the energy of the O-2$p$ feature at $\sim2$ eV binding energy, while the position of this feature is off by $\sim1$ eV in the $U$ = 9 eV calculation. Considering  the Ni $t_{2g}$ peak position, the $U$ = 7 eV calculation correctly reproduces this, while the $U$ = 5 and 9 eV calculations place this peak at  too low and high binding energies, respectively. As a result, only the $U$ = 7 eV calculation is able to correctly reproduce the energy scale of both features observed in experiment.

Now constraining $U$ to 7 eV, we consider the impact of varying $J$ in Fig.~\ref{fig:1}(b), and show the extracted O-2$p$ onset and Ni-$t_{2g}$ peak locations as a function of $J$ in Fig.~\ref{fig:1}(d). We find that the $J$ = 1.4 eV calculation captures the energy of the O-2$p$ onset, while the $J$ = 0.6 eV calculation places this feature at too low binding energy. The Ni $t_{2g}$ peak is located at too high and low binding energies, respectively, in the $J$ = 0.6 eV and 1.4 eV calculations. From consideration of these DFT + DMFT calculations with five different ($U$, $J$) parameter sets, we find that only ($U$, $J$) = (7, 1) eV is able to reproduce the experimental energy position of both features, while the other calculations misalign either one or both features. As a result we use this parameter set in the rest of our calculations. In addition to constraining the free parameters in our calculation, this comparison in Fig.~\ref{fig:1} also highlights the sensitivity of the energies of features in the angle-integrated photoemission spectrum to the electron interaction parameters used  in DFT + DMFT calculations. A previous study~\cite{barman} on bulk LaNiO$_3$ extracted a value of $U$ = 4.7 eV by comparing XPS and Auger spectra, which is lower than our optimized value of $U$. However, the extraction of U from XPS/Auger spectra is indirect, requiring modeling of core-hole excitonic corrections and hybridization effects, which may be why this value is different from that obtained from our photoemission/DMFT comparison.

\subsection{Reference bandstructure}
\begin{figure}
\includegraphics[width=0.45\textwidth]{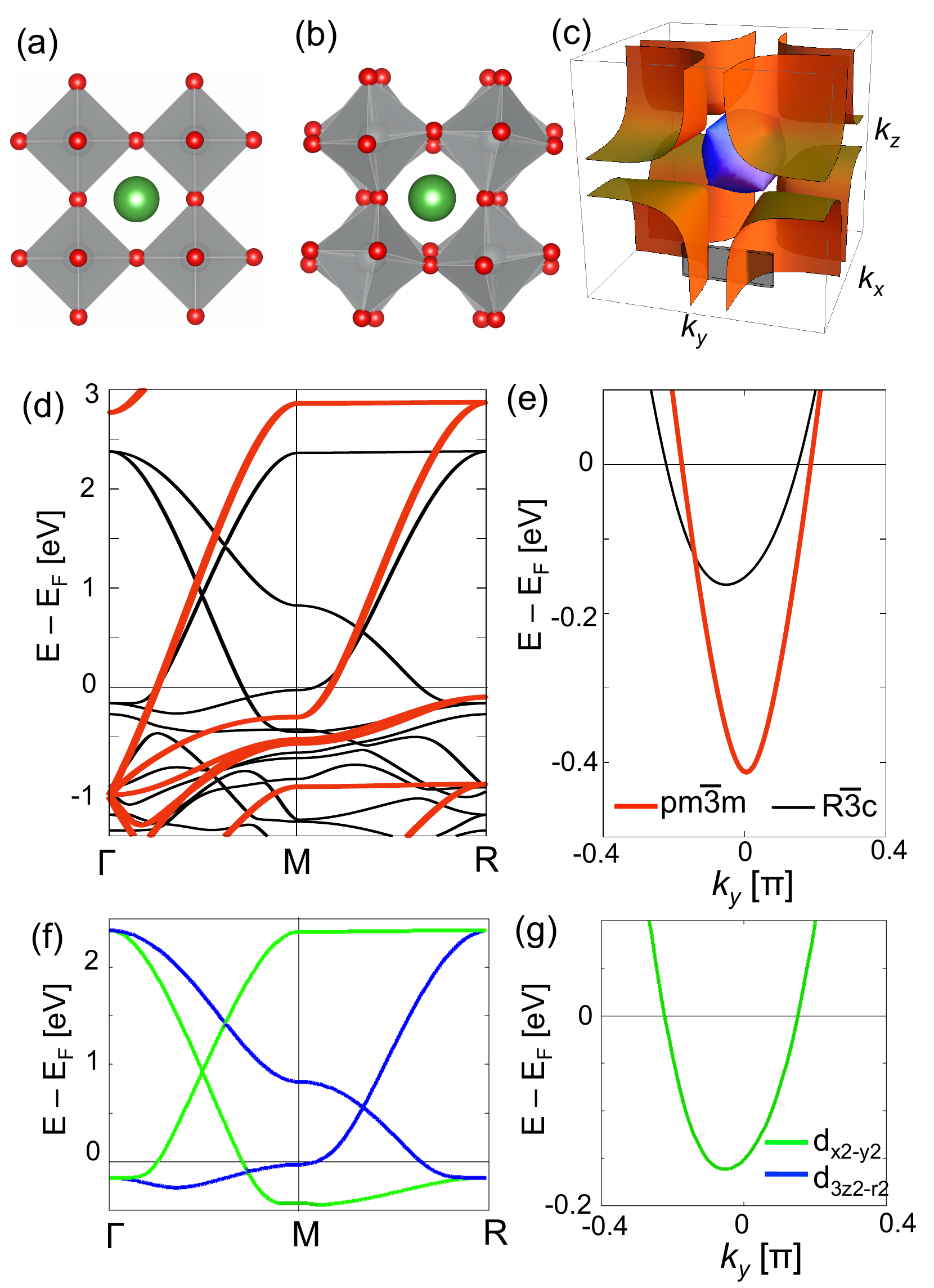}
\caption{\label{fig:2} { Bulk DFT LaNiO$_3$ bandstructure in cubic and rhombohedral structures.} Comparison of cubic (a)  and rhombohedral (b) structures: in the rhombohedral structure, the NiO$_6$ octahedra rotate out of phase about each of the pseudocubic axes. (c)  Fermi surface for LaNiO$_3$ in the cubic structure. The black slab indicates the momentum space cut ($\pi$/2$a_{pc}$, $k_y$, 0.7$\pi/a_{pc}$) that we measure in our ARPES experiment.  (d) and (e) compare the  cubic and rhombohedral bandstructures (thick red and thin black lines) along high-symmetry cuts and along the  experimental cut, respectively. Note that in (d) we label high-symmetry points using notation for the cubic structure. (f) and (g) show the dominant orbital character ($d_{x^2-y^2}$ or $d_{3z^2-r^2}$) of the bands crossing the Fermi level (for the rhombohedral structure) for the same momentum space cuts as shown in (d) and (e).
}
\end{figure}

We now consider the choice of  reference bandstructure. LaNiO$_3$ exhibits a rhombohedral structural distortion,~\cite{may} which has been observed experimentally in our films via superstructure in low-energy electron diffraction measurements.~\cite{king} We perform DFT  calculations for  two structures: a hypothetical cubic structure (shown in Fig.~\ref{fig:2}(a)) and the experimental rhombohedral ($R\bar{3}c$) structure, where   the NiO$_6$ octahedra rotate out of phase about the $[111]$ pseudocubic axis, which corresponds to rotations of equal amounts about each of the three pseudocubic axes, as shown in Fig.~\ref{fig:2}(b).  In each case we fully relax the structures; we find a  Ni-O-Ni bond angle of 161.6$^\circ$ and a Ni-O bond length of 1.95 $\AA$ when we relax the rhombohedral structure with PBE. This compares to  experimental values of 164.8$^\circ$ and 1.93 $\AA$ for the bond angle and length, respectively, for bulk LaNiO$_3$ at 1.5 K.~\cite{torrance}

Fig.~\ref{fig:2}(c) shows a schematic of the LaNiO$_3$ Fermi surface computed for the hypothetical cubic structure. It consists of two sheets with a small electron pocket centered at the $\Gamma$ point, and large hole  pockets centered at the Brillouin zone corners. In this work we study the bandstructure along the momentum space cut ($\pi$/2$a_{pc}$, $k_y$, 0.7$\pi/a_{pc}$), which is shown as the black slab in the lower part of Fig.~\ref{fig:2}(c) (this momentum space cut is determined by our experimental photon energy). We determine the value of $k_z$ corresponding to this photon energy as described in Ref.~\onlinecite{king}. Due to surface sensitivity, the photoemission measurements integrate over a range of $k_z$, which is represented by the finite height of  the slab along $k_z$ in Fig.~\ref{fig:2}(c). However, the electronic structure has little $k_z$-dependence in this range, so this should minimally affect our results. We also consider calculated bandstructures along high-symmetry cuts through the Brillouin zone for comparison.  LaNiO$_3$ has a nominal  $t_{2g}^6e_g^1$ electronic configuration, so the bands crossing the Fermi level have predominately $e_g$ character with a sizeable O $2p$ component from hybridization. The  dominant orbital character of the bands crossing the Fermi level is shown in Fig.~\ref{fig:2}(f-g).

The rhombohedral distortion present in LaNiO$_3$ is generally expected to reduce the bandwidth relative to that of the cubic structure, because rotations of the NiO$_6$ octahedra distort the Ni-O-Ni bond angle away from 180$^\circ$ and thus reduce the orbital overlap between the Ni $e_g$ and the O $2p$ states. To clarify how the octahedral rotations influence the specifics of the near-Fermi level bandstructure needed  for obtaining $m^\star/m$ from ARPES measurements, we compare DFT bandstructures  computed in the rhombohedral structure and in the idealized cubic structure both along high-symmetry Brillouin zone cuts (Fig.~\ref{fig:2}(d)) and along our experimentally accessible momentum cut (Fig.~\ref{fig:2}(e)). 

Focusing first on the bandstructure along high-symmetry cuts in Fig.~\ref{fig:2}(d), the two prominent near-Fermi level features are a band crossing the Fermi level on the $\Gamma$-$M$ cut (two bands in the case of the rhombohedral bandstructure, due to zone folding), and a shallow band bottom at the $M$ point. The Fermi velocity $v_F$ of the $\Gamma$-$M$ band is reduced  in the rhombohedral bandstructure relative to that in the cubic bandstructure, and the $M$-point band bottom moves to significantly lower binding energy. Both of these changes are consistent with the reduction of Ni-O hybridization due to the rhombohedral distortion. However, the magnitude of these changes differs substantially: while there is only a small difference in $v_F$ along $\Gamma$-$M$ between the rhombohedral and cubic structures, the position of the band bottom at the $M$ point moves from $\sim300$ meV binding energy (cubic) to $\sim30$ meV binding energy (rhombohedral). 

Now turning to the momentum space cut accessed by our experiments (Fig.~\ref{fig:2}(e)), we find that both $v_F$ and the binding energy of the band bottom are reduced by approximately a factor of 2 in the rhombohedral structure relative to the cubic one. (Note that the slight offset of the rhombohedral band relative to the cubic one is due to the lower symmetry of the rhombohedral structure). In summary, the comparison in Fig.~\ref{fig:2}(d-e) illustrates that the rhombohedral distortion present in LaNiO$_3$ has a significant influence on the low-energy bandstructure, but the precise magnitude of this effect varies significantly  in momentum space. Therefore, a quantitative determination of $m^\star/m$ depends sensitively on the correct  details of the crystal structure, and requires a comparison of theory and experiment at the same locations in momentum space.

\section{Results\label{sec:results}}
\subsection{Bulk-like rhombohedral structure \label{sec:rhom}}
\begin{figure}
\includegraphics[width=0.3\textwidth]{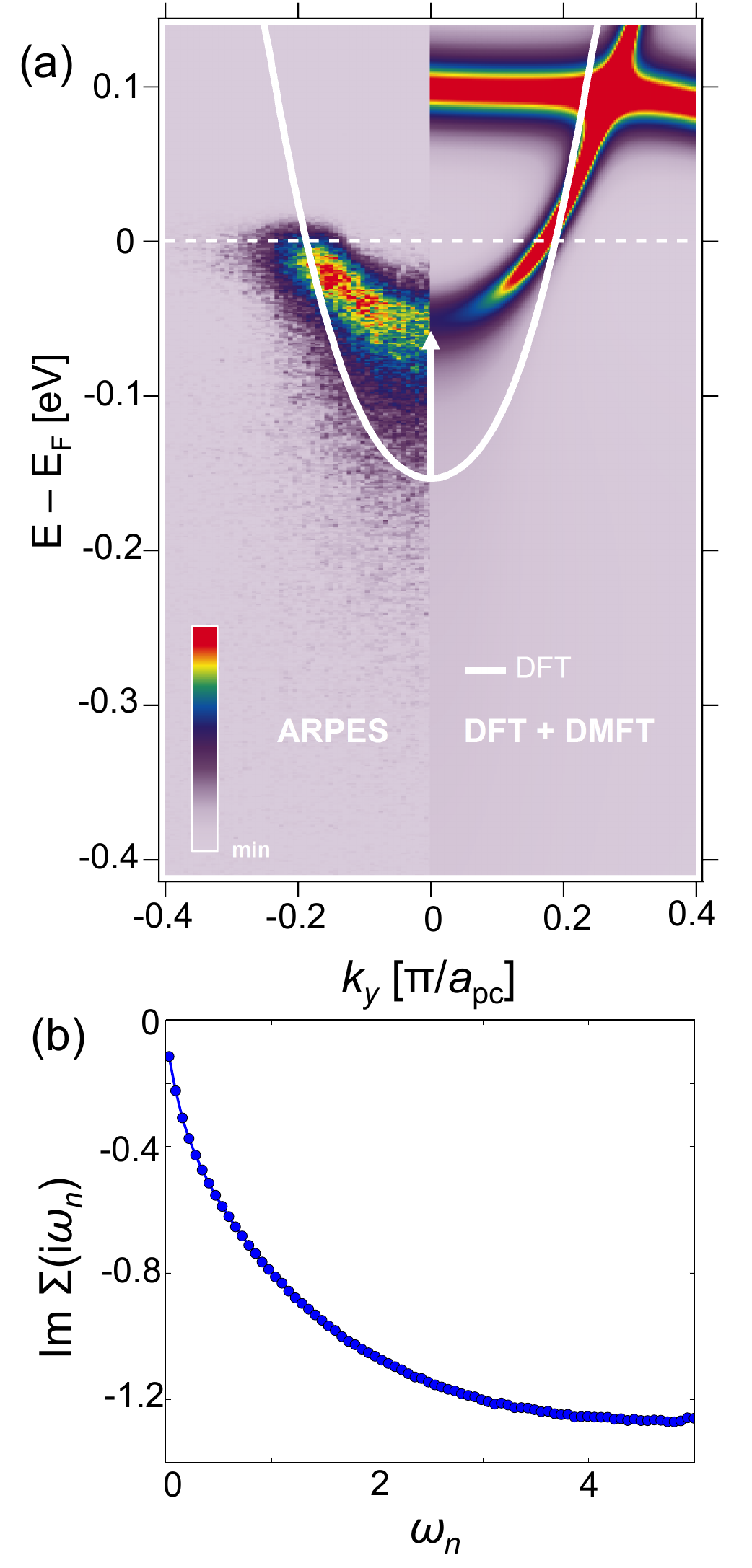}
\caption{\label{fig:3}  (a) Comparison of ARPES spectrum (left side) and DFT + DMFT spectral function (right side), both along the momentum space cut ($\pi$/2$a_{pc}$, $k_y$, 0.7$\pi /a_{pc}$) to the DFT bandstructure, calculated in the bulk $R\bar{3}c$ structure (white line).  (b) Imaginary part of the Matsubara axis DFT + DMFT self-energy, used to calculate the theoretical value of $m^\star/m$.
}
\end{figure}

Fig.~\ref{fig:3}(a) compares the rhombohedral DFT bandstructure from Fig.~\ref{fig:2}(e) to  spectra measured by ARPES and calculated by DFT + DMFT. The experiment and calculations consider  exactly the same momentum cut ($\pi$/2$a_{pc}$, $k_y$, 0.7$\pi/a_{pc}$).  In both experiment (left side) and DFT+DMFT  (right side), there is a shallow band crossing the Fermi level with a band bottom at $\sim$ 50 meV and a Fermi level crossing at $k_y=-0.2\pi/a_{pc}$.  This band is substantially renormalized by electron correlations relative to the rhombohedral DFT bandstructure, and the renormalization predicted by DFT+DMFT is in good agreement with that seen in experiment. It is important to emphasize that the DFT + DMFT calculation used the values of $U$ and $J$   obtained from matching features in the high-energy spectrum in Fig.~\ref{fig:1}, we do not further optimize these parameters to obtain the present comparison of the low-energy spectra. 
 
We obtain a mass renormalization of $m^\star/m = 3.1 \pm 0.5$ by comparing the DFT and experimental / DFT+DMFT   band bottom energies, and $m^\star/m = 3.4 \pm 0.5$ by comparing Fermi velocities $v_F$. An earlier report of $m^\star/m$ of 7 in this system from some of the present authors arose from using DFT calculations in the idealized cubic structure~\cite{king} which,  as can be seen from  Fig.~\ref{fig:2}(e),  are about a factor of two more dispersive. 

Alternatively, the theoretical mass renormalization can be obtained by considering the frequency derivative of the electron self-energy $\partial \Sigma' (\omega)/\partial \omega|_{\omega=0}$ (in the single-site DMFT approximation considered here,  the self-energy has no ${\bf k}$-dependence). This is related to the physical mass enhancement discussed in the previous paragraph by factors of the relative $d$ and $p$ content of the near-Fermi surface wave functions \cite{Wang2009,Wang2011}.  In the Fermi liquid regime relevant here,   the imaginary part of the real-axis self-energy is negligible at low frequencies and  one may estimate $\partial\Sigma'(\omega)/\partial \omega|_{\omega=0}$ from the imaginary part of the self-energy on the Matsubara axis $\Sigma^{\prime\prime}(i\omega_n)$, shown in Fig.~\ref{fig:3}(b). We fit the five lowest Matsubara points to a fourth order polynomial to obtain  $m^\star/m = (1-\partial \Sigma^{\prime\prime}(i\omega_n)/\omega_n |_{\omega_n \rightarrow 0}) = 3.5$. We also calculate the mass renormalization from the analytically continued self-energy, $m^\star/m = (1-\partial\Sigma^\prime(\omega)/\partial\omega |_{\omega = 0})$ (not shown) and obtain the same value, thus lending confidence to our analytic continuation procedure.  The estimate we obtain is in good agreement with the experimentally determined value, and consistent with the observation that in this region of the Brillouin zone the states are of primarily Ni-$d$ character.  This quantitative comparison between experiment and theory in Fig.~\ref{fig:3}(a) demonstrates that DFT + DMFT is able to accurately describe correlated physics in LaNiO$_3$. However, it also highlights the importance of considering realistic crystal structure, correct interaction parameters,  and precisely the same momentum space points in experiment and theory.

\subsection{Strained films}
\begin{figure}
\includegraphics[width=0.5\textwidth]{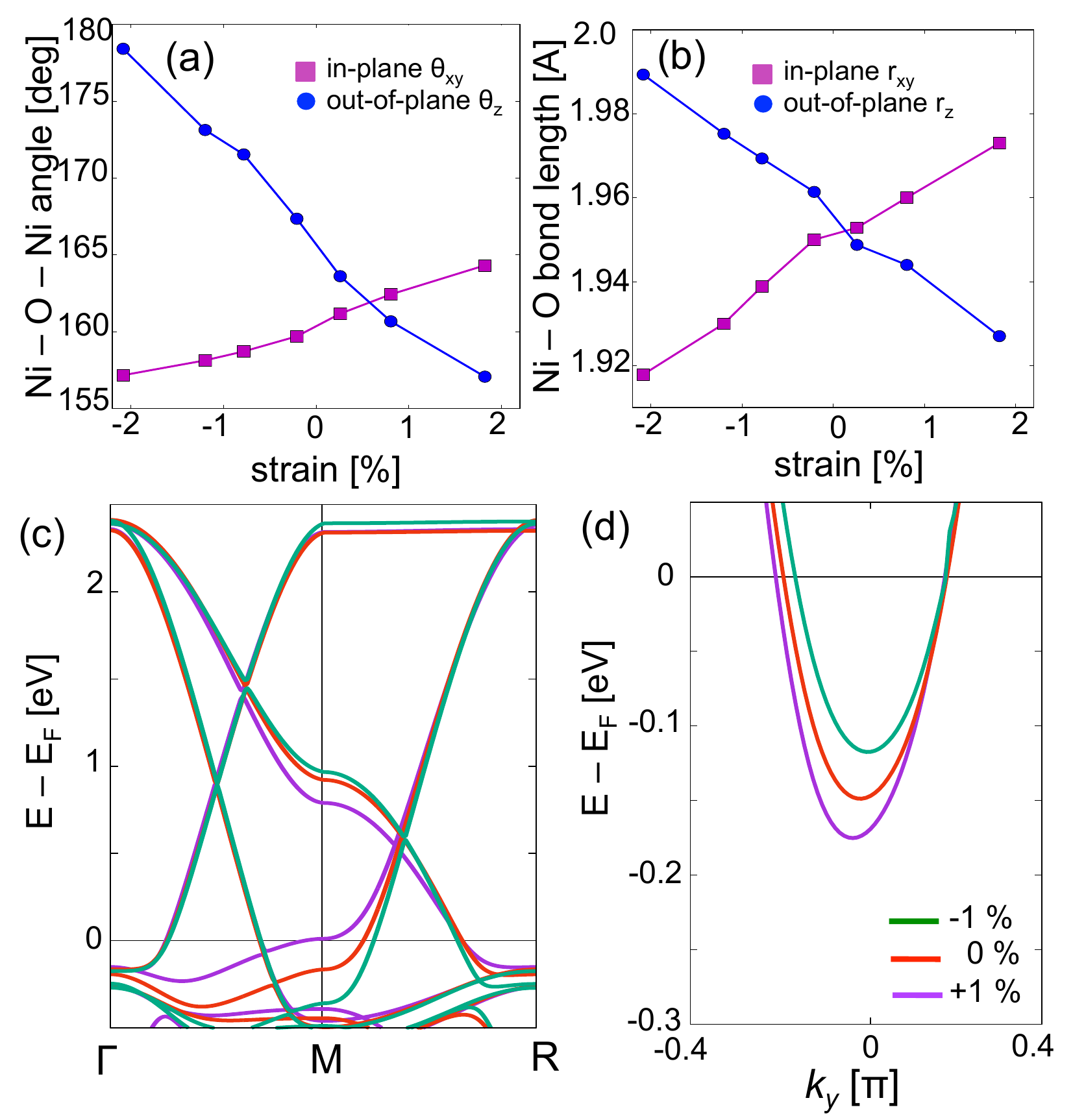}
\caption{\label{fig:4} {Influence of biaxial strain on DFT bandstructure.}  The evolution with strain of  (a) the Ni-O-Ni bond angles and (b) the Ni-O bond lengths that lie in (purple) and out (blue) of the plane in which biaxial strain is applied.  Here negative (positive) values correspond to compressive (tensile) strain.  The bandstructure at -1, 0, and +1$\%$ strain is shown in  (c) along high symmetry cuts, and (d) along the experimental ARPES cut. Note that LaNiO$_3$ grown on LaAlO$_3$ (the experimental setup) corresponds to 1$\%$ compressive strain.
}
\end{figure}

LaNiO$_3$ grown on LaAlO$_3$ is under $\sim1 \%$ compressive strain, so it is important to consider how this strain impacts our results, given that strain couples to octahedral rotations. Here  we define strain as $(a-a_0)/a_0$, where $a_0$ is the in-plane lattice constant at which the total energy of the $C2/c$ structure is a minimum, and apply strain in the $xy$ plane.  As discussed in Sec.~\ref{sec:methods}, the structure becomes monoclinic ($C2/c$) under the epitaxial constraint, which changes the pattern of octahedral rotations. The NiO$_6$ octahedra still rotate out of phase about [111]$_{pc}$, but there are now two distinct Ni-O-Ni bond angles $\theta_{xy}$ and $\theta_z$, which lie in and out of the plane of applied strain, respectively. In addition, there are also two distinct Ni-O bond lengths, $r_{xy}$ and $r_z$ lying in (out of) the strain plane. The crystal structure accommodates this biaxial strain via changes to both the Ni-O-Ni bond angles and the Ni-O bond lengths, as shown in Fig.~\ref{fig:4}(a-b).  Upon moving from compressive to tensile strain, $\theta_{xy}$  moves closer to 180$^\circ$ and $r_{xy}$ expands to accommodate the stretching of the crystal in this plane, while $\theta_z$  moves further from 180$^\circ$ and $r_z$  contracts.

Experimentally, an asymmetry in the response of the in-/out-of plane  bond angles and lengths is observed in LaNiO$_3$: the out-of-plane bond angle changes more with biaxial strain than the in-plane bond angle, while the in-plane bond length changes more than the out-of-plane bond length~\cite{may} (the same asymmetric response also is observed in strained LaAlO$_3$~\cite{johnson-wilke}). We find that our calculated bond angles in Fig.~\ref{fig:4}(a) reproduce this asymmetric trend, while the bond lengths in Fig.~\ref{fig:4}(b) do not ($r_{xy}$ and $r_z$ change by roughly the same amount). Improved agreement with experiment regarding these structural changes can be obtained by performing structural relaxations within DFT + $U$.~\cite{may}  

We compare the bandstructure at -1, 0, and +1$\%$ biaxial strain along high symmetry cuts and along our experimental cut in Fig.~\ref{fig:4}(c) and (d). Focusing on the near-Fermi level bandstructure features, in Fig.~\ref{fig:4}(c), $v_F$ of the band crossing the Fermi level on the $\Gamma$-$M$ cut is essentially unchanged for the strains we consider, while the position of the near-Fermi level band bottom at the $M$-point changes substantially (note that the 0$\%$-strain $C2/c$ bandstructure does not need to agree precisely with the bulk $R\bar{3}c$ bandstructure due to the different symmetries of these structures). For our experimental momentum cut in Fig.~\ref{fig:4}(d), the band displays moderate changes with strain. Interestingly, the response of the bandstructure to strain is remarkably different at different momentum points: compressive strain pushes the  the band bottom at the $M$ point  to higher binding energy in Fig.~\ref{fig:4}(c), while the band bottom in Fig.~\ref{fig:4}(d) moves to lower binding energy. 
These differences can be understood in light of the fact that biaxial strain lifts the degeneracy of the $e_g$ orbitals, and bands derived from the $d_{3z^2-r^2}$ and $d_{x^2-y^2}$ orbitals  respond differently to biaxial strain (both in terms of changes to the bandwidth and the band's center of mass).~\cite{peil}  As can be seen in Fig.~\ref{fig:2}(f-g), the band bottom at $M$ is predominantly of $d_{3z^2-r^2}$ character, while the band on our experimental cut is predominantly $d_{x^2-y^2}$ character, so it is not surprising that these bands display  opposite trends with strain.

While these results reveal a complex evolution of the bandstructure with strain, the key observation for our chosen system of LaNiO$_3$/LaAlO$_3$ is that in Fig.~\ref{fig:4}(d)  the band under $1\%$ compressive strain is quite similar to the bulk $R\bar{3}c$ band used to obtain $m^\star/m$ in Fig.~\ref{fig:3} (both in terms of band bottom energy and $v_F$). Therefore our use of the bulk $R\bar{3}c$ structure rather than the $1\%$ compressively strained $C2/c$ structure for the particular strain and momentum space cut considered in this work does not introduce significant errors in our determination of $m^\star/m$. However, as illustrated in Fig.~\ref{fig:4}, the bands at other strain values and momentum space cuts can be significantly different from the bulk bandstructure, so in general  using a bandstructure computed at the strain imposed by the experimental substrate may be necessary to obtain a quantitative comparison between experiment and theory.

\begin{figure}
\includegraphics[width=0.4\textwidth]{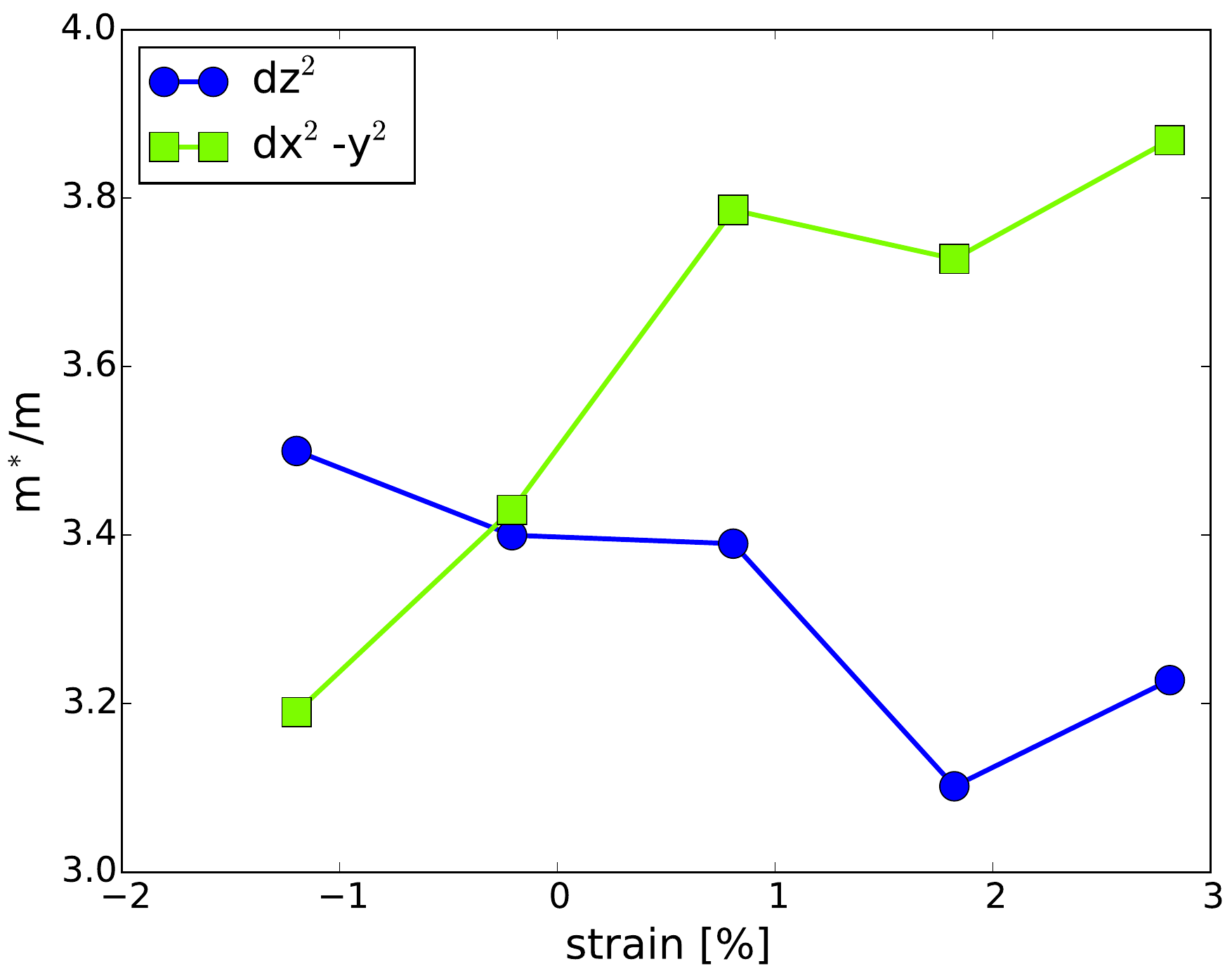}
\caption{\label{fig:5} { Strain-dependence of calculated DFT+DMFT mass enhancement} in the $a^-a^-c^-$ structure.  Because biaxial strain breaks the degeneracy of the $e_g$ orbitals, we show here orbitally-resolved mass enhancements for $d_{x^2-y^2}$  (green) and $d_{3z^2-r^2}$ (blue).
}
\end{figure}

Finally, Fig.~\ref{fig:5} shows the strain-dependence of the mass renormalization $m^\star/m$ calculated with DFT + DMFT. These values of $m^\star/m$ were obtained from fitting  Matsubara self-energies, using the same  procedure as in Sec.~\ref{sec:rhom}. Because biaxial strain breaks the degeneracy of the $e_g$ orbitals, there are now distinct self-energies corresponding to the $d_{x^2-y^2}$ and $d_{3z^2-r^2}$ orbitals, so Fig.~\ref{fig:5} shows the orbitally-resolved mass enhancements.  We find that the $d_{x^2-y^2}$  mass enhancement increases upon stretching  the in-plane lattice constant (moving from compressive to tensile strain), while the $d_{3z^2-r^2}$ mass enhancement  decreases slightly. Because the dominant orbital character of the bands crossing the Fermi level varies in momentum space, as shown in Fig.~\ref{fig:2}(f-g), at large compressive or tensile strains we would therefore expect a momentum-dependent mass enhancement if  ARPES measurements were made  at points around the Fermi surface with different dominant orbital characters.

The strain dependence of the orbitally-resolved $m^\star/m$ can be understood in light of the changes to the Ni-O-Ni bond angles and Ni-O bond lengths presented in Fig.~\ref{fig:4}(a-b), which both couple to the electronic bandwidth $W$.  Assuming strain does not strongly influence the electron-electron interaction strength $U$, strain-induced bandwidth changes will tune the ratio $U/W$, which controls the correlation strength.~\cite{imada}  Thus the increase (decrease) in the $d_{x^2-y^2}$ ($d_{3z^2-r^2}$)-orbitally resolved $m^\star/m$ with tensile strain would arise from a decrease (increase)  in bandwidth $W_{x^2-y^2}$ ($W_{3z^2-r^2}$).   For bands formed by hybridized $e_g$ and O $2p$ orbitals, $W$ is related to the Ni-O bond length $r$ and the Ni-O-Ni bond angle $\theta$ via the expression~\cite{harrison} $W  \sim |\cos\theta|/r^{3.5}$. Here,  the bandwidth  $W_{x^2-y^2}$ is controlled by $r_{xy}$ and $\theta_{xy}$, while  the bandwidth $W_{3z^2-r^2}$ is controlled by $r_z$ and $\theta_z$. 

The expression  for $W$ above reveals that  strain-induced changes to $r$ and $\theta$ shown in Fig.~\ref{fig:4}(a-b) will generally affect the bandwidth in opposite ways. Taking the example of  $W_{x^2-y^2}$, tensile strain increases $r_{xy}$ and $\theta_{xy}$ relative to the unstrained values, which will decrease (increase) the bandwidth, respectively.  As a result, the net bandwidth response to strain depends on whether changes in bond angle or bond length are the primary mechanism for strain accommodation, and to what extent the changes cancel each other out. 
Comparing Fig.~\ref{fig:4}(a-b) and Fig.~\ref{fig:5}, it is clear that the changes to the orbitally resolved $m^\star/m$ with strain follow the trends that one would expect from the Ni-O bond distance changes, rather than the Ni-O-Ni bond angle  (an increasing $d_{x^2-y^2}$ mass enhancement arises from increased $r_{xy}$, while a decreasing $d_{3z^2-r^2}$ $m^\star/m$ arises from a decreasing $r_z$). The fact that the $d_{x^2-y^2}$ mass enhancement changes more with strain than the $d_{3z^2-r^2}$ mass enhancement can be understood because $\theta_{xy}$ responds less to strain than $\theta_{z}$ as shown in Fig.~\ref{fig:4}(a), so there is less cancellation of the bond length/angle changes to the bandwidth.

While this analysis describes the strain induced changes to $m^\star/m$ solely in terms of bond angle and bond length changes, the evolution of correlation strength with strain may be more complex, due to the strong coupling between electronic correlations and the lattice in the $R$NiO$_3$ family.   For example, structural relaxations within DFT+$U$ ~\cite{may,chakhalian, blanca-romero} predict that tensile strain will induce a bond length disproportionation phase in LaNiO$_3$, similar to the correlation-induced disproportionation observed in the bulk insulating $R$NiO$_3$ compounds.~\cite{alonso,medarde}  
However, DFT+$U$  generally overpredicts the tendency towards bond disproportionation (while DFT underpredicts it), and structural relaxations within DFT + DMFT are needed to correctly reproduce the structural and pressure phase diagram of the bulk nickelates.~\cite{park2} On the experimental side, there is mixed evidence regarding bond disproportionation in strained LaNiO$_3$ films~\cite{chakhalian,Wu2013}. 

\section{Discussion\label{sec:lit}}
To summarize the last sections, comparison of ARPES  spectra to the DFT bandstructure computed in the rhombohedral structure  establishes a mass enhancement $m^\star/m\approx 3-3.5$ in LaNiO$_3$, and DFT+DMFT using the same structure, and interaction parameters fixed by high binding energy features in the photoemission spectra, give quasiparticle bands in very good agreement with the data.  This establishes LaNiO$_3$ as a moderately correlated Fermi liquid material. For context, this value of $m^\star/m$ is in the same regime as other correlated metallic oxides, such as SrRuO$_3$ ($m^\star/m \sim 4$)~\cite{alexander,shai} and VO$_2$ ($m^\star/m \sim 2$)~\cite{qazilbash}. 

Other measurements of  $m^\star/m$ for LaNiO$_3$  have previously appeared in the literature. Soft x-ray ARPES measurements~\cite{Eguchi} reported $m^\star/m \sim 3$ and $\sim 0$ on the electron- and hole-like bands, respectively, by comparing experiment to DFT calculations in the idealized cubic structure (as noted earlier, we obtain a renormalization of 7 when comparing our data to cubic DFT calculations). This difference of results could be due to measuring at different momentum space points, or to  the energy resolution available with soft x-rays. 

Optical conductivity experiments~\cite{Ouellette,stewart} determine a correlation strength from the integral of the optical conductivity over a given frequency range~\cite{Baeriswyl88,Gros86,Millis92}. One group~\cite{Ouellette} found an optical mass enhancement $m^\star/m\sim 3$  for LaNiO$_3$ grown on LaAlO$_3$, in good agreement with our results. They also reported that $m^\star/m$ increases from $\sim 3$ to $\sim 5$  as the strain moves from compressive to tensile. As seen in Fig.~\ref{fig:5} this trend is consistent with the results of  our DFT+DMFT calculations, but the experimental variation is larger than the calculated one.   Other optical conductivity experiments~\cite{stewart} found the same strain-dependent trend, but with larger mass enhancements.  The differences, both between experiments and between experiment and calculation, could arise partly from the choice of frequency range used in the analysis of optical data.  

In bulk  LaNiO$_3$, thermodynamic studies~\cite{xu,sreedhar,rajeev} report a Fermi surface averaged mass enhancement of $10$ relative to free electrons at a density corresponding to one electron per Ni site. Here we compare the experimental~\cite{rajeev} specific heat-based DOS at the Fermi level of $g(E_F) = (1.1-1.3) \times 10^{23}$ eV$^{-1}$ cm$^{-3}$ to the DOS obtained from our DFT calculations in the rhombohedral structure, $g_{DFT}(E_F) = 4.1\times 10^{22}$ eV$^{-1}$ cm$^{-3}$. Taking the ratio of the experimental and DFT DOS, we obtain $m^\star/m \sim g(E_F)/g_{DFT}(E_F)$ = 2.7 - 3.2, in reasonable agreement with the value we obtain from ARPES and DFT + DMFT in this paper. (For comparison, for the idealized cubic structure, we obtain $g_{DFT}(E_F) = 2.1\times 10^{22}$ eV$^{-1}$ cm$^{-3}$ and thus $m^\star/m$ = 5.2 - 6.2). The enhancement of the Fermi level DOS in the rhombohedral structure relative to the cubic structure is due to the flattening of the bands in the vicinity of the Fermi level in the rhombohedral structure as is visible in Fig.~\ref{fig:2}(d); this was previously discussed in Ref.~\onlinecite{peil}.

\section{Conclusion\label{sec:conclusion}}
In this paper, using ARPES measurements and DFT + DMFT calculations to study LaNiO$_3$, we determine the experimental and theoretical mass enhancement $m^\star/m$, which is a defining property of any correlated material, and demonstrate a quantitative comparison between experiment and theory. This result establishes LaNiO$_3$ as a moderately correlated Fermi liquid, and that DFT + DMFT can accurately describe this correlated physics. We compare our value of $m^\star/m$ to previous reports in the literature, and discuss possible origins of differences where appropriate; in particular, we find that our results agree with those from thermodynamic measurements, if we compare those measurements to our DFT calculations.

We highlight the choices that must be made in  such a experiment-theory comparison, in particular, the free parameters $U$ and $J$ in the DFT + DMFT calculation, which we constrain using angle-integrated photoemission spectra. We also emphasize the key role played by the reference DFT bandstructure.  Octahedral rotations and biaxial strain can change the near-Fermi level  bandstructure features significantly, and there is substantial variation in the magnitude of these changes depending on the particular point in momentum space under consideration. In general, octahedral rotations and strain can ``renormalize" the low-energy bands relative to an idealized cubic bandstructure by the same amount as electron correlations can, so it is key to consider the realistic crystal structure and biaxial strain (and make comparisons at the same momentum-space point) to obtain a quantitative estimate of the mass enhancement arising from electron correlations. These results provide an important benchmark for the DFT + DMFT method, and thus will enable future studies of other nickelates and strongly correlated materials.

\section{Acknowledgements} This work was supported by the Cornell Center for Materials Research with funding from the NSF MRSEC program (DMR-1120296) and the Office of Naval Research (N00014-12-1-0791).   J. P. R. acknowledges support from the NSF IGERT program (DGE-0903653). A. J. M. acknowledges support from the Basic Energy Sciences division of the Department of Energy under grant number ER-046169. H. P.  gratefully acknowledges the support of start-up funds from University of Illinois at Chicago and Argonne National Laboratory. Part of the computational work was carried out at computing facilities supported by the Cornell Center for Materials Research. H. P. acknowledges the computing resources provided by Blues, a high-performance computing cluster operated by the Laboratory Computing Resource Center at Argonne National Laboratory. 

\bibliographystyle{apsrev}
\bibliography{bib_nowadnick_LaNiO3}

\end{document}